# Applications of the Nilpotent Dirac State Vector

**Peter Rowlands\* and J. P. Cullerne†**

*\*IQ Group and Department of Physics, University of Liverpool, Oliver Lodge Laboratory, Oxford Street, P.O. Box 147, Liverpool, L69 7ZE, UK. e-mail prowl@hep.ph.liv.ac.uk and prowl@csc.liv.uk*

*†IQ Group, Department of Computer Science, University of Liverpool, Chadwick Laboratory, Peach Street, Liverpool, L69 72F, UK.*

*Abstract.* The nilpotent version of the Dirac equation is applied to the baryon wavefunction, the strong interaction potential, electroweak mixing, and Dirac and Klein-Gordon propagators. The results are used to interpret a quaternion-vector model of particle structures.

## 1 Introduction

The primary purpose of this work is physical, rather than mathematical. The Dirac algebra, in our interpretation, has a *physical*, rather than mathematical, origin. We have shown, in previous work,[1] that it is possible to derive a version of the Dirac equation from physical first principles which is at once equivalent to, but more powerful than, the standard version, incorporating a complete treatment of the quantum field, and replacing the more restricted matrix representation by a more general algebraic one. However, this new form of the equation is really *physically equivalent* to the standard version of Dirac, rather than algebraically isomorphic to it, and a literal application of the more restricted algebra used in the standard version will not produce it. Though we will sometimes use the term 'wavefunction' for our quaternion state vectors, these new mathematical objects have a much wider range of application than conventional wavefunctions, and their *physical* origin will lead to important applications in particle physics, which are not available via the standard interpretation.

Essentially, we use a 32-part algebra produced by combining the 4-vector units (*i*, **i**, **j**, **k**) with the quaternion units (1, *i*, *j*, *k*), where the two sets of terms correspond to the units of the fundamental parameters space-time and mass-charge. Then, beginning with the relativistic energy conservation equation,

$$E^2 - p^2 - m^2 = 0 ,$$

we first factorize and attach the exponential term $e^{-i(Et - \mathbf{p}\cdot\mathbf{r})}$, so that

$$(\pm k E \pm i\mathbf{i}\, \mathbf{p} + i\mathbf{j}\, m)\, (\pm k E \pm i\mathbf{i}\, \mathbf{p} + i\mathbf{j}\, m)\, e^{-i(Et - \mathbf{p}\cdot\mathbf{r})} = 0 .$$

Then replacing $E$ and $\mathbf{p}$ in the first bracket with the quantum operators, $i\partial / \partial t$ and $-i\nabla$, to give

$$\left(\pm ik\frac{\partial}{\partial t} \pm i\nabla + ijm\right)(\pm kE \pm ii\,\mathbf{p} + ij\,m)\,e^{-i(Et - \mathbf{p.r})} = 0\,,$$

we obtain a quantum mechanical equation of the form the form

$$\left(\pm ik\frac{\partial}{\partial t} \pm i\nabla + ijm\right)\psi = 0\,,$$

where the wavefunction (or, in our terminology, the quaternion state vector),

$$\psi = (\pm kE \pm ii\,\mathbf{p} + ij\,m)\,e^{-i(Et - \mathbf{p.r})}\,.$$

The most convenient form of this operator, of course, arranges the four terms which it incorporates in a column (or ket) vector, while the four terms in the differential operator take up equivalent positions in a row (or bra) vector. The differential operator (or left-hand bracket) may be then considered as introducing the variable part of the equation, while the state vector operator (or right-hand bracket) specifies the conserved part.

**2 The quaternion state vector (QSV)**

The Dirac equation overspecifies its components. The algebra, as normally written, specifies the same information 3 times: in the $E$-$\mathbf{p}$-$m$ terms, in the spinors, and in the exponentials. In the nilpotent formulation, the information is specified only once, in the first term of the quaternion state vector (QSV). The QSV then automatically selects the 3 remaining terms in sequence, incorporating all values of $\pm E \pm \mathbf{p}$. The exponential term for a free particle is algebraically the same for any state, and determined only by the values of $E$ and $\mathbf{p}$. The same principle applies for a bound state, though here the $E$ and $\mathbf{p}$ terms are determined as eigenvalues produced by the differential operator acting on the variable or nonquaternionic part of the wavefunction. The Hamiltonian is completely determined by the QSV, the differential operator having an eigenvalue identical to the state vector, when operating on the common variable term. The nature of the variable term is totally determined by whatever function is necessary to produce the correct nilpotent when acted on by the differential operator. The anticommuting pentad term

$$(kE + ii\mathbf{p} + ij\,m)$$

completely defines a state of a free fermion, spin up, because the variable term is necessarily always $e^{-i(Et - \mathbf{p.r})}$, for *any* free state, and the complete specification of the state vector follows automatically, as:

$$\begin{pmatrix} kE + i i \mathbf{p} + ijm \\ kE - i i \mathbf{p} + ijm \\ -kE + i i \mathbf{p} + ijm \\ -kE - i i \mathbf{p} + ijm \end{pmatrix}$$

The differential operator is identical to the state vector, but with the *E* and **p** representing the relevant quantum operators with or without field terms, rather than eigenvalues. The first term of the state vector codifies *all* the information about a state. The exponential term in the free particle case is an aspect of the automatic application of the state to the vacuum. In defining states, including composite ones, we don't need to use the variable term at all. We simply use vector multiplication of the quaternion state vectors.

**3 Vacuum operator**

The vacuum operator applied simultaneously to all four solutions is most conveniently represented by a diagonal matrix, premultiplied by a row state vector or postmultiplied by a column state vector. In the first case, we write:

$$( (-kE - i i \mathbf{p} + ijm) \; (-kE + i i \mathbf{p} + ijm) \; (kE - i i \mathbf{p} + ijm) \; (kE + i i \mathbf{p} + ijm) ) \times$$

$$k \begin{pmatrix} kE + i i \mathbf{p} + ijm & 0 & 0 & 0 \\ 0 & kE - i i \mathbf{p} + ijm & 0 & 0 \\ 0 & 0 & -kE + i i \mathbf{p} + ijm & 0 \\ 0 & 0 & 0 & -kE - i i \mathbf{p} + ijm \end{pmatrix} e^{-i(Et - \mathbf{p}.\mathbf{r})}$$

$$= ( (-kE - i i \mathbf{p} + ijm) \; (-kE + i i \mathbf{p} + ijm) \; (kE - i i \mathbf{p} + ijm) \; (kE + i i \mathbf{p} + ijm) ) \, e^{-i(Et - \mathbf{p}.\mathbf{r})} \,,$$

assuming the appropriate normalisation constants. The vacuum wavefunction operator (when applied to a row vector) is always $k \times$ matrix form of state vector $\times$ exponential term. The vacuum operator omits the exponential term. The order is reversed when applied to a column vector.

$$\begin{pmatrix} kE + ii\mathbf{p} + ijm & 0 & 0 & 0 \\ 0 & kE - ii\mathbf{p} + ijm & 0 & 0 \\ 0 & 0 & -kE + ii\mathbf{p} + ijm & 0 \\ 0 & 0 & 0 & -kE - ii\mathbf{p} + ijm \end{pmatrix} k$$

$$\times \begin{pmatrix} kE + ii\mathbf{p} + ijm \\ kE - ii\mathbf{p} + ijm \\ -kE + ii\mathbf{p} + ijm \\ -kE - ii\mathbf{p} + ijm \end{pmatrix} e^{-i(Et - \mathbf{p}.\mathbf{r})} = \begin{pmatrix} kE + ii\mathbf{p} + ijm \\ kE - ii\mathbf{p} + ijm \\ -kE + ii\mathbf{p} + ijm \\ -kE - ii\mathbf{p} + ijm \end{pmatrix} e^{-i(Et - \mathbf{p}.\mathbf{r})}$$

The process may be repeated indefinitely in each case without alteration to the fermion state.

**4 The hydrogen atom**

The derivation of the hyperfine structure for the hydrogen atom may be taken as a useful test of the power of any Dirac formalism. Using the nilpotent QSV, we can reduce the procedure to relatively simple algebra applied to a single equation. In principle, we have begin with the nilpotent equation

$$(\pm kE \pm ii\, \mathbf{p} + ij\, m)\,(\pm kE \pm ii\, \mathbf{p} + ij\, m) = 0 \;.$$

As a product of row and column vectors, this can be written:

$$((kE + ii\mathbf{p} + ijm)\;(kE - ii\mathbf{p} + ijm)\;(-kE + ii\mathbf{p} + ijm)\;(-kE - ii\mathbf{p} + ijm)) \begin{pmatrix} kE + ii\mathbf{p} + ijm \\ kE - ii\mathbf{p} + ijm \\ -kE + ii\mathbf{p} + ijm \\ -kE - ii\mathbf{p} + ijm \end{pmatrix} = 0 \quad (1)$$

For a free fermion, this leads to the Dirac equation:

$$\left(\pm ik\frac{\partial}{\partial t} \pm i\boldsymbol{\sigma}.\nabla + ijm\right)(\pm kE \pm ii\mathbf{p} + ijm)\, e^{-i(Et - \mathbf{p}.\mathbf{r})} = 0 \;,$$

where the differential operator is understood to be a row vector, and the wavefunction a column vector, with the same four components as in (1).

If we now apply a potential $\phi$, to a fermion of charge $-e$,

$$i\frac{\partial}{\partial t} \rightarrow i\frac{\partial}{\partial t} + e\phi,$$

or, for eigenvalue $E$,

$$E \rightarrow E + e\phi.$$

The Dirac equation now takes the form

$$(\pm \mathbf{k}(E + e\phi) \pm i\mathbf{\sigma}.\nabla + i\mathbf{j}m)\,\psi = 0.$$

The $\pm$ values of $\mathbf{k}$ and $i$ still lead to four solutions within $\psi$. In the case of stationary states, this implies that $\psi$ contains a nilpotent column vector of the form $(\pm \mathbf{k}E' \pm i\mathbf{i}\mathbf{p}' + i\mathbf{j}m)$, where $E'$ and $\mathbf{p}'$ are terms with the respective dimensions of energy and momentum.

For the hydrogen atom, we assume a potential $\phi$ of the form $Ze/r$, and write the expression $\mathbf{\sigma}.\nabla$ as a function of $r$ in polar coordinates, with an explicit angular momentum term for the electron:

$$\mathbf{\sigma}.\nabla = \left(\frac{\partial}{\partial r} + \frac{1}{r}\right) \pm i\frac{j + \frac{1}{2}}{r}$$

An explicit angular momentum term is not required where $\nabla$ can be regarded as a multivariate vector, but, here, $\nabla$ is regarded, for the convenience of explicitly incorporating the electron's angular momentum states, as an ordinary vector. We therefore write the Dirac equation for the electron in a central potential as:

$$\left(\mathbf{k}\left(E + \frac{Ze^2}{r}\right) + i\left(\frac{\partial}{\partial r} + \frac{1}{r} \pm i\frac{j + \frac{1}{2}}{r}\right) + i\mathbf{j}m\right)\psi = 0.$$

Suppose the variable part of $\psi$ has the form

$$F = e^{-ar}\,r^\gamma \sum_{\nu = 0} a_\nu r^\nu.$$

Then

$$\frac{\partial F}{\partial r} = \left(-a + \frac{\gamma}{r} + \frac{\nu}{r} + \ldots\right) F,$$

where, for a bound state, $a$ is real and positive. The nilpotent part of the wavefunction (for stationary states) must be of the exact form to zero the eigenvalues produced by the differential operator:

$$\left(\pm \mathbf{k}\left(E + \frac{Ze^2}{r}\right) \pm i\left(-a + \frac{\gamma}{r} + \frac{\nu}{r} + \ldots + \frac{1}{r} \pm i\frac{j + \frac{1}{2}}{r}\right) + i\mathbf{j}m\right).$$

In column vector form, this becomes:

$$\begin{pmatrix} \left(k\left(E + \dfrac{Ze^2}{r}\right) + i\left(-a + \dfrac{\gamma}{r} + \dfrac{v}{r} + \ldots + \dfrac{1}{r} + i\dfrac{j + \frac{1}{2}}{r}\right) + ijm\right) \\ \left(k\left(E + \dfrac{Ze^2}{r}\right) - i\left(-a + \dfrac{\gamma}{r} + \dfrac{v}{r} + \ldots + \dfrac{1}{r} - i\dfrac{j + \frac{1}{2}}{r}\right) + ijm\right) \\ \left(-k\left(E + \dfrac{Ze^2}{r}\right) + i\left(-a + \dfrac{\gamma}{r} + \dfrac{v}{r} + \ldots + \dfrac{1}{r} + i\dfrac{j + \frac{1}{2}}{r}\right) + ijm\right) \\ \left(-k\left(E + \dfrac{Ze^2}{r}\right) - i\left(-a + \dfrac{\gamma}{r} + \dfrac{v}{r} + \ldots + \dfrac{1}{r} - i\dfrac{j + \frac{1}{2}}{r}\right) + ijm\right) \end{pmatrix}$$

If we take the eigenvalues of the differential operator as a row vector of exactly the same form, and equate the product of the row and column vectors to zero, we find that:

$$4\left(E + \dfrac{Ze^2}{r}\right)^2 = -2\left(-a + \dfrac{\gamma}{r} + \dfrac{v}{r} + \dfrac{1}{r} + i\dfrac{j + \frac{1}{2}}{r}\right)^2 - 2\left(-a + \dfrac{\gamma}{r} + \dfrac{v}{r} + \dfrac{1}{r} - i\dfrac{j + \frac{1}{2}}{r}\right)^2 + 4m^2.$$

Equating constant terms, we find

$$E^2 = -a^2 + m^2,$$

$$a = \sqrt{m^2 - E^2}$$

Equating terms in $1/r^2$, with $v = 0$, we obtain:

$$\left(\dfrac{Ze^2}{r}\right)^2 = -\left(\dfrac{\gamma + 1}{r}\right)^2 + \left(\dfrac{j + \frac{1}{2}}{r}\right)^2,$$

from which, excluding the negative root (as usual),

$$\gamma = -1 + \sqrt{(j + \tfrac{1}{2})^2 - (Ze^2)^2}.$$

Assuming the power series terminates at $n'$, and equating coefficients of $1/r$ for $v = n'$,

$$2EZe = -2\sqrt{m^2 - E^2}\,(\gamma + 1 + n'),$$

the terms in $(j + \tfrac{1}{2})$ cancelling over the summation of the four multiplications. From this we may derive

$$\dfrac{E}{m} = \left(1 + \dfrac{(Ze^2)^2}{(\gamma + 1 + n')^2}\right)^{-1/2},$$

or

$$\dfrac{E}{m} = \left(1 + \dfrac{(Ze^2)^2}{(\sqrt{(j + \frac{1}{2})^2 - (Ze^2)^2} + n')^2}\right)^{-1/2}.$$

This, with $Z = 1$, is the fine structure formula for the hydrogen atom.

## 5 Baryons

Baryon wavefunctions may also be constructed from the nilpotents, using the three-dimensional properties of the **p** operator. Suppose we represent the six degrees of freedom for the spin as $\pm p_1, \pm p_2, \pm p_3$. Then an expression of the form

$$(k\mathbf{E} \pm i\mathbf{i}\, p_1 + i\mathbf{j}\, m)(k\mathbf{E} \pm i\mathbf{i}\, p_2 + i\mathbf{j}\, m)(k\mathbf{E} \pm i\mathbf{i}\, p_3 + i\mathbf{j}\, m) \qquad (2)$$

has the same structure as the fermionic $(k\mathbf{E} + i\mathbf{i}\,\mathbf{p} + i\mathbf{j}\, m)$ when we equate **p** successively with $\pm p_1, \pm p_2, \pm p_3$. We can then incorporate this as the first term in a row or column vector, with four solutions, and treat it the same as any other fermion.

The standard QCD representation of the baryon is the antisymmetric colour singlet of $SU(3)$:
$$\psi \sim (BGR - BRG + GRB - GBR + RBG - RGB).$$

Here, we use a mapping such as:

$$
\begin{array}{ll}
BGR & (k\mathbf{E} + i\mathbf{j}\, m)(k\mathbf{E} + i\mathbf{j}\, m)(k\mathbf{E} + i\mathbf{i}\,\mathbf{p} + i\mathbf{j}\, m) \\
-BRG & (k\mathbf{E} + i\mathbf{j}\, m)(k\mathbf{E} - i\mathbf{i}\,\mathbf{p} + i\mathbf{j}\, m)(k\mathbf{E} + i\mathbf{j}\, m) \\
GRB & (k\mathbf{E} + i\mathbf{j}\, m)(k\mathbf{E} + i\mathbf{i}\,\mathbf{p} + i\mathbf{j}\, m)(k\mathbf{E} + i\mathbf{j}\, m) \\
-GBR & (k\mathbf{E} + i\mathbf{j}\, m)(k\mathbf{E} + i\mathbf{j}\, m)(k\mathbf{E} - i\mathbf{i}\,\mathbf{p} + i\mathbf{j}\, m) \\
RBG & (k\mathbf{E} + i\mathbf{i}\,\mathbf{p} + i\mathbf{j}\, m)(k\mathbf{E} + i\mathbf{j}\, m)(k\mathbf{E} + i\mathbf{j}\, m) \\
-RGB & (k\mathbf{E} - i\mathbf{i}\,\mathbf{p} + i\mathbf{j}\, m)(k\mathbf{E} + i\mathbf{j}\, m)(k\mathbf{E} + i\mathbf{j}\, m), \qquad (3)
\end{array}
$$

with each term equivalent to $-p^2(k\mathbf{E} + i\mathbf{i}\,\mathbf{p} + i\mathbf{j}\, m)$ or $-p^2(k\mathbf{E} - i\mathbf{i}\,\mathbf{p} + i\mathbf{j}\, m)$. This gives the same three cyclic and three anticyclic combinations as the conventional representation. Because there is only one spin term, it also predicts that the spin is a property of the baryon wavefunction as a whole, not of component quark wavefunctions.

This structure is determined solely by the nilpotent nature of the fermion wavefunction. Put in an extra **p** into the brackets missing them, and we immediately reduce to zero. With the spinor terms included, each of these is represented by a tensor product of three spinors, for example:

$$(k\mathbf{E} + i\mathbf{j}\, m)(k\mathbf{E} + i\mathbf{j}\, m)(k\mathbf{E} + i\mathbf{i}\,\mathbf{p} + i\mathbf{j}\, m)\left(\frac{1}{2}\right) \otimes \left(\frac{1}{2}\right) \otimes \left(\frac{1}{2}\right)$$

where

$$\left(\frac{1}{2}\right) \otimes \left(\frac{1}{2}\right) \otimes \left(\frac{1}{2}\right) = \left(\frac{3}{2}\right) \oplus \left(\frac{1}{2}\right) \oplus \left(\frac{1}{2}\right)$$

So this representation encompasses both spin ½ and spin 3/2 baryon states.

The baryon structure is presumed to be maintained by a strong interaction between the three component (quark) states, maintained by an exchange of massless gluons. The *SU*(3) symmetry for this strong source is conventionally expressed using a covariant derivative of the form:

$$\partial_\mu \to \partial_\mu + ig_s \frac{\lambda^\alpha}{2} A^{\alpha\mu}(x).$$

In component form:

$$ip_1 = \partial_1 \to \partial_1 + ig_s \frac{\lambda^\alpha}{2} A^{\alpha 1}(x)$$

$$ip_2 = \partial_2 \to \partial_2 + ig_s \frac{\lambda^\alpha}{2} A^{\alpha 2}(x)$$

$$ip_3 = \partial_3 \to \partial_3 + ig_s \frac{\lambda^\alpha}{2} A^{\alpha 3}(x)$$

$$E = i\partial_0 \to i\partial_0 - g_s \frac{\lambda^\alpha}{2} A^{\alpha 0}(x).$$

Using this, we may observe that the baryon state vector has the same form as the eigenvalue of the Dirac differential operator, which is the product of the three terms:

$$\left(k\left(E - g_s \frac{\lambda^\alpha}{2} A^{\alpha 0}\right) \pm i\left(\partial_1 + ig_s \frac{\lambda^\alpha}{2} A^{\alpha 1}\right) + ij\,m\right)$$

$$\left(k\left(E - g_s \frac{\lambda^\alpha}{2} A^{\alpha 0}\right) \pm i\left(\partial_2 + ig_s \frac{\lambda^\alpha}{2} A^{\alpha 2}\right) + ij\,m\right)$$

$$\left(k\left(E - g_s \frac{\lambda^\alpha}{2} A^{\alpha 0}\right) \pm i\left(\partial_3 + ig_s \frac{\lambda^\alpha}{2} A^{\alpha 3}\right) + ij\,m\right).$$

Because of the underlying nilpotent nature of the term ($kE \pm i\mathbf{i}\,\mathbf{p} + i\mathbf{j}\,m$), the only way of preserving nonzero fermionic structure here is to write this expression in one of the forms:

$$\left(k\left(E - g_s \frac{\lambda^\alpha}{2} A^{\alpha 0}\right) \pm i\left(\partial_1 + ig_s \frac{\lambda^\alpha}{2} \mathbf{A}^\alpha\right) + ij\,m\right)\left(k\left(E - g_s \frac{\lambda^\alpha}{2} A^{\alpha 0}\right) + ij\,m\right)\left(k\left(E - g_s \frac{\lambda^\alpha}{2} A^{\alpha 0}\right) + ij\,m\right)$$

$$\left(k\left(E - g_s \frac{\lambda^\alpha}{2} A^{\alpha 0}\right) + ij\,m\right)\left(k\left(E - g_s \frac{\lambda^\alpha}{2} A^{\alpha 0}\right) \pm i\left(\partial_1 + ig_s \frac{\lambda^\alpha}{2} \mathbf{A}^\alpha\right) + ij\,m\right)\left(k\left(E - g_s \frac{\lambda^\alpha}{2} A^{\alpha 0}\right) + ij\,m\right)$$

$$\left(k\left(E - g_s \frac{\lambda^\alpha}{2} A^{\alpha 0}\right) + ij\,m\right)\left(k\left(E - g_s \frac{\lambda^\alpha}{2} A^{\alpha 0}\right) + ij\,m\right)\left(k\left(E - g_s \frac{\lambda^\alpha}{2} A^{\alpha 0}\right) \pm i\left(\partial_1 + ig_s \frac{\lambda^\alpha}{2} \mathbf{A}^\alpha\right) + ij\,m\right),$$

which are, of course, parallel to the six forms expressed in (2) and (3).

In effect, this means that the carrier of the 'colour' component of the strong force ($ig_s\,\lambda^\alpha\,\mathbf{A}^\alpha / 2$) is 'transferred' between the quarks at the same time as the spin, both being incorporated into the **p** term, and the current that effects the 'transfer' is carried by the gluons or generators of the strong field. This is the physical meaning of the transfer of the charge component *s* between representations A, B and C (or between the three 'colours' R, G and B) in the charge accommodation tables (for which, see section 9). To make the baryon wavefunction noncollapsable, of course, and the strong interaction gauge invariant, all the representations or 'phases' are present at the same time, and equally probable. We could say that the three quark 'colours' are no more capable of separation from each other than are the three dimensions of space (and, indeed, these conditions are exactly equivalent to each other, and determine the nature of the strong force). So the idea of a 'transfer' of strong charge or 'colour' field is, in effect, simply a convenient way of expressing the innate gauge invariance of the strong interaction, at the same time as conservation of angular momentum.

**6 The quark-antiquark and three-quark interactions**

Lattice gauge calculations from QCD suggest that the quark-antiquark potential in the bound meson state at the quenched (long-distance) level is, at least approximately, of the form[3]:

$$V = -\frac{A}{r} + \sigma r + C,$$

as required also by experimental studies of charmonium states. Here, for convenience, we abbreviate $V_{Q\bar{Q}}$, $A_{Q\bar{Q}}$, $\sigma_{Q\bar{Q}}$, $C_{Q\bar{Q}}$, by $V$, $A$, $\sigma$ and $C$. Assuming a strong or colour charge for the quark of strength $q$ ($= \sqrt{\alpha_s}$), we have a potential energy

$$W = q\frac{A}{r} - q\sigma r - qC$$

for the interactions of quark and antiquark ($qA$ being worked out theoretically at $4\alpha_s / 3$). (It will be significant that the constant term $C$ has no effect on the form of the results obtained, merely shifting the value of $E$ to $E - qC$.) We can now construct the nilpotent operator, in the same manner as that for the hydrogen atom:

$$\mathbf{k}\left(E + q\frac{A}{r} - q\sigma r - qC\right) + \mathbf{i}\left(\frac{\partial}{\partial r} + \frac{1}{r} \pm i\frac{j + \tfrac{1}{2}}{r}\right) + \mathbf{ij}m.$$

For convenience, this is shown as a single term but the complete operator requires a column vector incorporating the four possible combinations of $\pm \mathbf{k}$ and $\pm \mathbf{i}$. The $\pm i$ ($j + \tfrac{1}{2}$) term arises from the multivariate nature of the σ.∇ operation, in the same way as for the electron in the hydrogen atom. Two of the four Dirac solutions require positive values and two negative. Initially, we suppose that the nonquaternionic part of the wavefunction has the form

$$\psi = \exp(-ar - br^2)\, r^\gamma \sum_{v = 0} a_v r^v,$$

and consider the ground state (with $v = 0$) over the four Dirac solutions. The four-part nilpotent wavefunction defines the condition:

$$4\left(E + q\frac{A}{r} - q\sigma r - qC\right)^2 = -2\left(\frac{\partial}{\partial r} + \frac{1}{r} + i\frac{j + \tfrac{1}{2}}{r}\right)^2 - 2\left(\frac{\partial}{\partial r} + \frac{1}{r} - i\frac{j + \tfrac{1}{2}}{r}\right)^2 + 4m^2$$

for all solutions.

Applying $\psi$ and expanding, we obtain:

$$(E - qC)^2 - 2q^2A\sigma + \frac{q^2A^2}{r^2} + q^2\sigma^2 r^2 + \frac{2qA}{r}(E - qC) - 2q\sigma(E - qC)r$$

$$= -\left(a^2 + \frac{(\gamma + \nu + 1)^2}{r^2} - \frac{(j + \frac{1}{2})^2}{r^2} + 4b^2 r^2 + 4abr - 4b(\gamma + \nu + 1) - \frac{2a}{r}(\gamma + \nu + 1)\right) + m^2 .$$

The positive and negative $i(j + \frac{1}{2})$ terms cancel out over the four solutions as they do in the case of the hydrogen atom. We now equate:

(1) coefficients of $r^2$:

$$q^2\sigma^2 = -4b^2$$

(2) coefficients of $r$:

$$-2q\sigma(E - qC) = -4ab$$

(3) coefficients of $1/r$:

$$-2qA(E - qC) = 2a(\gamma + \nu + 1)$$

(4) coefficients of $1/r^2$:

$$q^2A^2 = -(\gamma + \nu + 1)^2 + (j + \frac{1}{2})^2$$

(5) constant terms:

$$(E - qC)^2 - 2q^2A\sigma = -a^2 + 4b(\gamma + \nu + 1) + m^2$$

From the first three equations, we immediately obtain:

$$b = \pm \frac{iq\sigma}{2}$$

$$a = \mp i(E - qC)$$

$$\gamma + \nu + 1 = \mp iqA .$$

The case where $\nu = 0$, then leads to

$$(j + \frac{1}{2})^2 = 0$$

$$m^2 = 0 .$$

This suggests a wavefunction with variable component

$$\psi = \exp(\mp i(E - qC)r \pm iq\sigma r^2/2) \, r^{\pm iqA - 1}$$

for the ground state, with $v = 0$. If we can assign physical meaning to the case where $v \neq 0$, and the power series in $\psi$ terminates in $v = n'$, we will conclude that

$$q\sigma = -i\frac{m^2}{2n'}$$

and

$$qA = -i\frac{(j + \tfrac{1}{2})^2 - n'^2}{2n'} ,$$

requiring the power series to be composed of negative imaginary integers.

The imaginary exponential terms in $\psi$ can be interpreted as representing asymptotic freedom, the $\exp(\mp i(E - qC)r$ being typical for a free fermion. The complex $\exp(\pm iq\sigma r^2/2)$ term is similar to the real one used for a harmonic oscillator. The $r^{\gamma-1}$ term is also complex, and can be written as a phase, $\phi(r) = \exp(\pm iqA \ln(r))$, which varies less rapidly with $r$ than the rest of $\psi$. We can therefore write $\psi$ in the form

$$\psi = \frac{\exp(kr + \phi(r))}{r} ,$$

where

$$k = (\mp i(E - qC) \pm iq\sigma r/2) .$$

When $r$ is small (at high energies), the first term dominates, approximating to a free fermion solution (which can be interpreted as asymptotic freedom). When $r$ is large (at low energies) the second term dominates, bringing in the confining potential ($\sigma$) (which can be interpreted as producing infrared slavery).

It is significant that no spherically symmetric solution can be reached, under any conditions, with a potential $\propto r$, without the additional Coulomb term, because the spherical symmetry introduces terms in $1/r$ and $1/r^2$ as coefficients of $i^2$ which must be negated by similar terms acting as coefficients of $k^2$. The algebraic structure of the nilpotent representation also rules out a confining potential proportional to $\ln r$. A confining potential proportional to $r$ implies a constant force, and, as the form of the solution remains unchanged by the presence of a constant term in the potential, the requirements for asymptotic freedom and infrared slavery are met simply by assuming that the quark confining force must be constant in magnitude and equal in all directions.

In line with theoretical expectations, we can show that, if the quark-quark potential is reduced to the Coulomb term, as might be imagined to happen effectively at short distances, we obtain a hydrogen-like spectral series. Here, we have

$$4\left(E + q\frac{A}{r} - qC\right)^2 = -2\left(\frac{\partial}{\partial r} + \frac{1}{r} + i\frac{j + \tfrac{1}{2}}{r}\right)^2 - 2\left(\frac{\partial}{\partial r} + \frac{1}{r} - i\frac{j + \tfrac{1}{2}}{r}\right)^2 + 4m^2 ,$$

where the nonquaternionic part of the wavefunction must have the form

$$\psi = \exp(-ar)\, r^\gamma \sum_{\nu=0} a_\nu r^\nu,$$

with the $\exp(-br^2)$ no longer required. On application of this function over the four Dirac solutions, and expansion (for the ground state), we obtain:

$$(E - qC)^2 + \frac{q^2 A^2}{r^2} + \frac{2qA}{r}(E - qC)$$

$$= -\left(a^2 + \frac{(\gamma + \nu + 1)^2}{r^2} - \frac{(j + \tfrac{1}{2})^2}{r^2} - \frac{2a}{r}(\gamma + \nu + 1)\right) + m^2.$$

This time, there are only three equations – for coefficients of $1/r$, coefficients of $1/r^2$, and constant terms:

$$2qA(E - qC) = 2a(\gamma + \nu + 1)$$

$$q^2 A^2 = -(\gamma + \nu + 1)^2 + (j + \tfrac{1}{2})^2$$

$$(E - qC)^2 = -a^2 + m^2,$$

leading to:

$$a = \frac{qA(E - qC)}{(\gamma + \nu + 1)}$$

$$(\gamma + \nu + 1) = \pm\sqrt{(j + \tfrac{1}{2})^2 - q^2 A^2}$$

$$m^2 = (E - qC)^2 \left(1 + \frac{q^2 A^2}{(\gamma + \nu + 1)^2}\right).$$

Significantly, below a certain value of $(E - qC)$, $a$ is real, suggesting a confined solution. Also, the status of $\gamma$ is determined by the values of $\nu$ and $j$; while $m$ here is nonzero. The equations are identical in form to those for the hydrogen atom with $qA$ replacing $Ze^2$, and $E - qC$ replacing $E$. We assume a wavefunction, with nonquaternionic component:

$$\phi\psi = \exp(-\sqrt{m^2 - (E - qC)^2})\, r^\gamma \sum_{\nu=0} a_\nu r^\nu,$$

and, allowing the power series to terminate at $\nu = n'$, we obtain the characteristic hydrogen-like solution:

$$\frac{E - qC}{m} = \left(1 + \frac{q^2 A^2}{(\gamma + 1 + n')^2}\right)^{-1/2},$$

or

$$\frac{E - qC}{m} = \left(1 + \frac{q^2 A^2}{(\sqrt{(j + \tfrac{1}{2})^2 - q^2 A^2} + n')^2}\right)^{-1/2}.$$

For a real system, such as charmonium, involving additional electrostatic terms, we can modify the Coulomb term by adding the appropriate electrostatic term (say $4e^2/9r$ or $e^2/9r$) to $qA/r$.

Rather than signifying escape, as with the electron in the hydrogen atom, the condition resulting from $(E - qC)^2 > m^2$ is that of *asymptotic* freedom, because of the continued presence (but reduced effect) of the confining linear potential. We can use the full and Coulomb-like solutions to make an approximate numerical calculation of the distance at which infrared slavery becomes effective.[4] From the full solution, let

$$k = (\mp i(E - qC) \pm iq\sigma r/2) = \frac{2\pi(r)}{\lambda},$$

and take $\lambda = \infty$ at zero energy (or infrared slavery). Then

$$q\sigma r = 2(E - qC)$$

and

$$r = \frac{2(E - qC)}{q\sigma}.$$

From the Coulomb-like solution, we take $(E - qC)$ as the mass or reduced mass of the c quark ($\approx 1.5$ GeV). Taking $\sigma \approx 1$ GeV fm$^{-1}$ and $q \approx 0.4$, we find $r \approx 4$ fm.

Virtually identical arguments apply to the three-quark or baryon system Here, the potential is of the form[5]:

$$V_{3Q} = -A_{3Q} \sum_{i<j} \frac{1}{|\mathbf{r}_i - \mathbf{r}_j|} + \sigma_{3Q} L_{min} + C_{3Q},$$

where $L_{min}$, the minimal total length of the colour flux tubes linking three quarks, arranged in a triangle with sides, $a$, $b$, $c$, is given by

$$L_{min} = \left[ \frac{1}{2}(a^2 + b^2 + c^2) + \frac{\sqrt{3}}{2}\sqrt{(a+b+c)(-a+b+c)(a-b+c)(a+b-c)} \right]^{1/2}.$$

For perfect spherical symmetry, when $a = b = c$, $L_{min}$ becomes a multiple of the distance $r$ of any quark from the centre of the flux tubes, and

$$\sum_{i<j} \frac{1}{|\mathbf{r}_i - \mathbf{r}_j|}$$

becomes a multiple of $1/r$. The potential $V_{3Q}$ then has exactly the same form as $V_{Q\bar{Q}}$, and the same solutions will apply, with variations in the values of the constants $A$, $\sigma$ and $C$. The model of Takahashi et al[5] suggests that $\sigma_{3Q} \approx \sigma_{Q\bar{Q}}$ and $A_{3Q} \approx A_{Q\bar{Q}}/2$, which accords with the theoretically-assumed value of $2\alpha_s/3$ for $qA$. It is highly likely that the relationship $A_{3Q} \approx A_{Q\bar{Q}}/2$ is virial in origin.

It is possible that the results may imply that quark mass ($m$) and intrinsic angular momentum ($j + \frac{1}{2}$) are both zero for quarks in the asymptotically free state, perhaps indicating that quark 'masses' might run to zero at that state, and to lepton masses at grand unification, and that the spins and masses of baryon states do not come from the individual valence quarks, but from the system as a whole, as both our baryon wavefunctions and charge accommodation rules imply, in addition to the results following on from the EMC experiment. The phase term in the full solution is interestingly proportional to $\alpha_s^2$, and is the only place where $A$ appears in the expression. Thus the Coulomb part of the potential – which is the component we believe to be significant in grand unification – results in a phase term (as does the $U(1)$ term for the electromagnetic interaction). It may be that we can regard this phase term as the one representing the gauge invariant 'transfer' of strong charge, or angular momentum, or vector part of the $SU(3)$ covariant derivative, between the 'coloured' components of baryons and mesons. It is, finally, this process of gauge invariant 'transfer' which allows us to suggest the derivation of the form of the confining potential from first principles, for the 'carrier' of the strong charge (or vector part of the covariant $SU(3)$ derivative) is the angular momentum, and constant force of equal magnitude in all directions is equivalent to a constant rate of change of an angular momentum, defined as $\sigma.\mathbf{p}$. The exact equivalence of all possible phases is identical to a constant rate of imagined rotational transfer of the strong charge via gluons in the same way as $c$ represents the constant rate of transfer of the electromagnetic force via virtual photons (and makes the relationship $\sigma_{3Q} \approx \sigma_{Q\bar{Q}}$ highly probable). From this fact alone, we can derive the necessity for a confining potential $\propto r$, which the nilpotent Dirac algebra requires to be supplemented by a Coulomb term representing the phase.

**7 The electroweak interaction**

Weak interactions all follow the same pattern. In the case of leptons, it is

$$e + \nu \rightarrow e + \nu. \qquad (4)$$

For quarks, it is

$$u + d \rightarrow u + d,$$

with d taking the place of $e$, and $u$ that of $\nu$. For weak interactions involving both leptons and quarks (for example, $\beta$ decay), the pattern is once again the same:

$$d + \nu \rightarrow e + u.$$

Let us, for the moment, consider (4). There are four possible vertices (assuming left-handed components only).

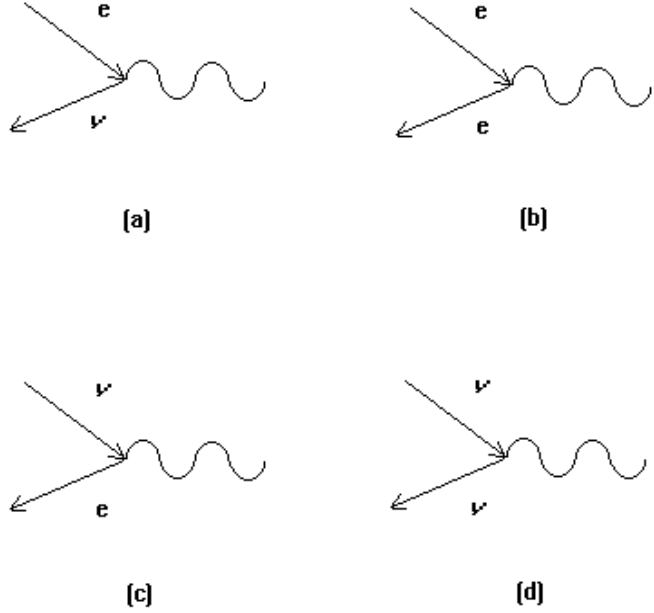

All the vertices must be true at once. The interaction is effectively a mixing or superposition of the four possibilities. However, the second vertex (b), and this one alone, also represents a possible electromagnetic interaction, giving us a 1 to 4 ratio for the occurrence of the electromagnetic to weak interaction at the energy which the vertices characteristically represent (that of the *W* / *Z* bosons). This means that particle charge structures at this energy must be such that the mixing ratio,

$$\sin^2 \theta_W = \frac{e^2}{w^2} = \frac{\Sigma t_3^2}{\Sigma Q^2} = 0.25 \ .$$

If we take the state vectors for the fermionic components of the four vertices, we obtain, for the case where the spins of the interacting fermions are assumed parallel (total 0 for fermion-antifermion combination):

(a) $(kE - ii\mathbf{p} + ij m) \ldots (-kE + ii\mathbf{p}) \ldots = 4m^2$ ;
(b) $(kE - ii\mathbf{p} + ij m) \ldots (-kE + ii\mathbf{p} + ij m) \ldots = 4m^2$ ;
(c) $(kE - ii\mathbf{p}) \ldots (-kE + ii\mathbf{p} + ij m) \ldots = 4m^2$ ;
(d) $(kE - ii\mathbf{p}) \ldots (-kE + ii\mathbf{p}) \ldots = 4m^2$ .

where $(kE - ii\mathbf{p} + ij m) \ldots$ represents a column or row vector with the terms:

$(kE - ii\mathbf{p} + ij m); (kE + ii\mathbf{p} + ij m); (-kE + ii\mathbf{p} + ij m); (-kE - ii\mathbf{p} + ij m)$ ,

and so on. Applying the usual normalisation, these sums become $m^2 / E^2$. Because (a)-(d) represent conditions of equal probability, then *E*, **p**, *m* must have identical values in all four cases. Hence, even if $\nu$ and $\bar{\nu}$ are massless objects, the $\nu\bar{\nu}$ vertex represented by (d) will not be massless. It is possible, however, that the $\nu\bar{\nu}$ combination is *intrinsically* massless (another 1 in 4 condition), and that the mass arises from mixing with the (b) vertex.

It is noticeable that, without an *m* term, all four vertices would become 0, and this *m* term arises from the fact that **p** is not purely composed of left-handed helicity states (with – **p** right-handed), but incorporates a right-handed component, which itself cannot contribute to the weak interaction because of charge-conjugation violation and the presence of a weak filled vacuum. The right-handed component *can only arise from the presence of the electromagnetic interaction*. In principle, we see that the weak interaction cannot exist as a pure left-handed interaction, without a mixing with the electromagnetic interaction to produce the necessary non-zero mass through the introduction of right-handed states. In particular, the purely weak $\nu\bar{\nu}$ (d) cannot exist independently of the other neutral vertex $e\bar{e}$ (b).

Suppose we put into the *E* and **p** terms of the state vector the covariant derivatives for the electroweak interaction. The scalar part goes with *E* and the vector part with **p**. Mass is produced by the mixing of *E* with **p** via the relativistic connection. It is also produced by the mixing of $B^0$ with $W^+$, $W^0$, and $W^-$, which we may now identify with the four vertices (d), (a), (b), and (c). By choosing the single, well-defined direction of spin or angular momentum (**p**) to be, in principle, the one where the total value for the interacting fermion-antifermion combination is 0, we can ensure that the mixing is specifically between the neutral components, $B^0$ and $W^0$, and create one massless *combination* to represent the carrier of the pure electromagnetic interaction ($\gamma$), with the other being the massive neutral weak carrier $Z^0$. The mixing must be such as to define the ratio of the two interactions, $\sin^2\theta_W$, at 0.25. (The other two vertices, $W^+$ and $W^-$, then fulfil the requirements for the existence of states corresponding to total spin values of +1 and –1.)

For left-handed leptons, we have the covariant derivatives:

$$\partial_\mu \to \partial_\mu + ig\frac{\tau.W^\mu}{2} - ig'\frac{B^\mu}{2} ,$$

and, for right-handed:

$$\partial_\mu \to \partial_\mu - ig'\frac{B^\mu}{2} .$$

Taking the energy operator and the single well-defined component of spin angular momentum, we have:

$$E = i\partial_0 \to i\partial_0 + g'\frac{B^0}{2} + ig'\frac{B^3}{2}$$

and

$$ip_3 = \partial_3 \to \partial_3 + ig\frac{\tau.W^3}{2} + ig\frac{\tau.W^0}{2} .$$

So, we can write the state vector for the (d) vertex in the form:

$$(k\mathbf{E} - i i\mathbf{p}) \ldots (-k\mathbf{E} + i i\mathbf{p}) \ldots = \left(k\left(\partial_0 + g'\frac{B^0}{2} + g'\frac{B^3}{2}\right) - i\left(\partial_3 + ig\frac{\tau.W^3}{2} + ig\frac{\tau.W^0}{2}\right)\right) \times$$

$$\left(-k\left(\partial_0 + g'\frac{B^0}{2} + g'\frac{B^3}{2}\right) + i\left(\partial_3 + ig\frac{\tau.W^3}{2} + ig\frac{\tau.W^0}{2}\right)\right)$$

and the state vector for the (b) vertex in the form:

$$(k\mathbf{E} - i i\mathbf{p} + ij m) \ldots (-k\mathbf{E} + i i\mathbf{p} + ij m) \ldots =$$

$$\left(k\left(\partial_0 + g'\frac{B^0}{2} + g'\frac{B^3}{2}\right) - i\left(\partial_3 + ig\frac{\tau.W^3}{2} + ig\frac{\tau.W^0}{2}\right) + ij m\right) \times$$

$$\left(-k\left(\partial_0 + g'\frac{B^0}{2} + g'\frac{B^3}{2}\right) + i\left(\partial_3 + ig\frac{\tau.W^3}{2} + ig\frac{\tau.W^0}{2}\right) + ij m\right).$$

Because *m* is determined from the combination of *E* and *p*, we can, by appropriate choice of the value of *m*, make these compatible if we additionally define a combination of *g'* and *g* which removes $B^3$ from *E* and $W^0$ from **p**. It is, of course, significant here that it is $B^\mu$ which is characteristic of right-handed lepton states, and therefore associated with the production of mass. Writing these combinations as $\gamma^0$ and $Z^3$, and those of *g'* and *g*, as *e* and *w* (= *g*), we obtain:

$$(k\mathbf{E} - i i\mathbf{p} + ij m) \ldots (-k\mathbf{E} + i i\mathbf{p} + ij m) \ldots =$$

$$\left(k\left(\partial_0 + e\frac{\gamma^0}{2}\right) - i\left(\partial_3 + iw\frac{\tau.Z^3}{2}\right) + ij m\right)\left(-k\left(\partial_0 + e\frac{\gamma^0}{2}\right) + i\left(\partial_3 + iw\frac{\tau.Z^3}{2}\right) + ij m\right).$$

Here, $\gamma^0 / 2$ becomes the same as the electrostatic potential $\phi$. So, we can write this in the form:

$$(k\mathbf{E} - i i\mathbf{p} + ij m) \ldots (-k\mathbf{E} + i i\mathbf{p} + ij m) \ldots =$$

$$\left(k(\partial_0 + e\phi) - i\left(\partial_3 + iw\frac{\tau.Z^3}{2}\right) + ij m\right)\left(-k(\partial_0 + e\phi) + i\left(\partial_3 + iw\frac{\tau.Z^3}{2}\right) + ij m\right).$$

Because *e* and *w* now represent the pure electromagnetic and weak coupling constants, we must necessarily obtain the ratio $e^2 / w^2 = 0.25$, and both quarks and leptons must be structured to observe this.

It should be noted here that *exchange* of electromagnetic charge, through, say, $W^+$ or $W^-$, is nothing to do with the electromagnetic interaction, but is rather an indication that the weak interaction is unable to detect the presence of the electromagnetic charge, that is, that a 'weak interaction' is a statement that all states of a particle with the same weak charge are equally probable, given the appropriate energy conditions, and that gauge invariance is maintained with respect to them. In principle, weak bosons are massive because they act as carriers of the electromagnetic charge, whereas electromagnetic bosons (or photons) are massless because they do not – the quantitative value of the mass must be determined from the coupling of the weak charge to the asymmetric vacuum state which produces the violation of charge conjugation in the weak interaction. The weak interaction is also indifferent to the presence of the strong charge, and so cannot distinguish between quarks and leptons – hence, the intrinsic identity of purely lepton weak interactions with quark-lepton or quark-quark ones – and, in the case of quarks, it cannot tell the difference between a filled 'electromagnetic vacuum' (up quark) and an empty one (down quark). The weak interaction, in addition, is also indifferent to the sign of the weak charge, and responds (via the vacuum) only to the status of fermion or antifermion – hence, the CKM mixing.

**8 Dirac and Klein-Gordon propagators**

We include this as an example of a result which demonstrates that the method we have employed is effective in many areas, even when the term ($kE + ii\mathbf{p} + ijm$) is no longer a nilpotent. It is also relevant to our work in sections 6 and 7 on the strong and electroweak interactions. In QED, we write the Dirac propagator

$$S_F(x - x') = (i\gamma^\mu \partial_\mu + m)\Delta_F(x - x'),$$

where $\Delta_F(x - x')$ is the Klein-Gordon propagator.[6] In our notation, we can write this in the form:

$$S_F(x - x') = ((kE + ii\mathbf{p} + ijm) \ldots)\Delta_F(x - x'),$$

where $((kE + ii\mathbf{p} + ijm) \ldots)$ is the bra matrix with the terms:

$$(kE + ii\mathbf{p} + ijm)$$
$$(kE - ii\mathbf{p} + ijm)$$
$$(-kE + ii\mathbf{p} + ijm)$$
$$(-kE - ii\mathbf{p} + ijm).$$

This is exactly what we would expect in transferring from boson (Klein-Gordon field) to fermion (Dirac field), using our single vector operator. Adapting the usual procedure, using the Green's function for the plane wave solutions, for the case in which variation over space and time (including the time-reversed solutions produced by $-E$ states) is transferred to the differential operator, we can then simply write

$$S_F(x - x') = \int d^3p \sqrt{\frac{m}{E}} \frac{1}{2E} (2\pi)^{-3/2} \theta(t - t') \Psi(x) \overline{\Psi}(x'),$$

where

$$\Psi(x) = ((kE + ii\mathbf{p} + ijm) \ldots) \exp(ipx)$$

and the adjoint term,

$$\overline{\Psi}(x') = ((kE - i\boldsymbol{i}\, \mathbf{p} - i\boldsymbol{j}\, m) \ldots) \exp(-ipx'),$$

with $((kE - i\boldsymbol{i}\, \mathbf{p} - i\boldsymbol{j}\, m) \ldots)(i\boldsymbol{k})$ now a ket. No averaging over spin states or 'interpreting' $-E$ as a reversed time state is necessary; the 'reversed time' state occurs with the $t$ in the operator $\partial / \partial t$. Reinterpreting $\Psi(x)$ and $\overline{\Psi}(x')$ as the vacuum expectation values of quantized spinor fields, say $\psi(x)$ and $\overline{\psi}(x')$, we obtain results of the form:

$$i\, S_F\, (x - x')_{ab} = \langle 0|\, T\, \psi(x)_a\, \overline{\psi}_b\, (x')\, |0\rangle.$$

In effect, multiplying bra terms of the form $(kE + i\boldsymbol{i}\, \mathbf{p} + i\boldsymbol{j}\, m) \ldots)$ with ket terms of the form $((kE - i\boldsymbol{i}\, \mathbf{p} - i\boldsymbol{j}\, m) \ldots)(i\boldsymbol{k})$ results in a scalar multiple of the bra term, while the exponential multiple takes the form $\exp(ip(x - x'))$.

**9 Charge accommodation**

In the usual approach to the Standard Model, the quantum fields have equations of motion generated by the Euler-Lagrange equations, using Lagrangians that possess the local gauge symmetries $SU(3)$ and $SU(2)_L \times U(1)$. In developing an interpretation of this model, rather than beginning with quantum fields, we have taken the three main conservation laws – conservation of baryon number, conservation of lepton number and conservation of electric charge – to have a fundamental origin, and have associated these laws with a parameter 'charge', which is three-dimensional.[7] The components of this parameter are identified with the sources of the three fundamental interactions: strong, weak and electromagnetic.

Now, the three conservation laws are separate and distinct, but what if at some energy these three types of charge were really equivalent? With this in mind we have searched for a way of assigning unit values of the three (at present) distinct types of charge to an underlying structure that makes no distinction between the three 'directions' we assign our unit values to. What makes this a plausible hypothesis is that it is impossible to make the three 'directions' indistinguishable unless unit charges only ever occur in combinations. As long as *all* possible combinations are simultaneously present, the charge types can remain distinct (that is, lead to distinct conservation laws), whilst the three 'directions', can remain indistinct with full exchange symmetry.

We have shown previously[7] that a value of the weak mixing parameter, $\sin^2\theta_W = 0.25$, at $\mu = M_Z$, would allow a Grand Unification at the Planck mass, in which the weak, strong and electromagnetic forces would have equal status in what may be a $U(5)$ structure incorporating gravity, and that such a suggestion could be tested experimentally by measuring the electromagnetic fine structure constant at higher energies – the electromagnetic $\alpha$, for example, increases to $1 / 118$ at 14 TeV. One way of generating this value of $\sin^2\theta_W = 0.25$ (which would be slightly reduced as an experimental value if measured through the production of real *W* and *Z* particles[8]) is to reconsider the Han-Nambu integrally-charged coloured quark theory,[9] which was largely discarded for general use (though not entirely in principle[10]) before the parallel phenomenon of the fractional quantum Hall effect was discovered and explained in condensed matter physics.[11] Laughlin, who explained the fractional quantum Hall effect as resulting from a single fermion forming a bosonic-type state with an odd number of magnetic flux lines, has recently hinted at its relevance for explaining fractional charges in particle physics,[12] and there seems to be no reason why, in a *fully* gauge invariant theory of the strong interaction, in which the quark colours are intrinsically inseparable, the underlying charges could not be integral while always being perceived as fractional in effect.[10]

We have investigated the possibilities of such a theory,[7] based on the idea that, while the separate weak, strong and electromagnetic charges may represented by quaternion labels, such as $\mathbf{k}, \mathbf{i}, \mathbf{j}$, the fundamental principle that each type of charge is conserved separately requires an extra vector-like degree of freedom to express the nonconservation or rotation symmetry of the quaternion operators which may be applied. Allowing for the fact that charges may come in zero or unit values, it has been possible to represent possible fermion states in two ways. One is a set of 'quark' tables A-E, which are shown, in reduced form, below:

**A**

|   |      | B   | G   | R   |
|---|------|-----|-----|-----|
| u | $+e$ | 1$j$ | 1$j$ | 0$i$ |
|   | $+s$ | 1$i$ | 0$k$ | 0$j$ |
|   | $+w$ | 1$k$ | 0$i$ | 0$k$ |
|   |      |     |     |     |
| d | $-e$ | 0$j$ | 0$k$ | 1$j$ |
|   | $+s$ | 1$i$ | 0$i$ | 0$k$ |
|   | $+w$ | 1$k$ | 0$j$ | 0$i$ |
|   |      |     |     |     |

**B**

|   |      | B   | G   | R   |
|---|------|-----|-----|-----|
| u | $+e$ | 1$j$ | 1$j$ | 0$k$ |
|   | $+s$ | 0$i$ | 0$k$ | 1$i$ |
|   | $+w$ | 1$k$ | 0$i$ | 0$j$ |
|   |      |     |     |     |
| d | $-e$ | 0$i$ | 0$k$ | 1$j$ |
|   | $+s$ | 0$j$ | 0$i$ | 1$i$ |
|   | $+w$ | 1$k$ | 0$j$ | 0$k$ |
|   |      |     |     |     |

**C**

|   |      | B   | G   | R   |
|---|------|-----|-----|-----|
| u | $+e$ | 1$j$ | 1$j$ | 0$k$ |
|   | $+s$ | 0$i$ | 1$i$ | 0$j$ |
|   | $+w$ | 1$k$ | 0$k$ | 0$i$ |
|   |      |     |     |     |
| d | $-e$ | 0$j$ | 0$k$ | 1$j$ |
|   | $+s$ | 0$i$ | 1$i$ | 0$k$ |
|   | $+w$ | 1$k$ | 0$j$ | 0$i$ |
|   |      |     |     |     |

**D**

|   |      | B   | G   | R   |
|---|------|-----|-----|-----|
| u | $+e$ | 1$j$ | 1$j$ | 0$i$ |
|   | $+s$ | 0$k$ | 1$i$ | 0$j$ |
|   | $+w$ | 0$i$ | 0$k$ | 1$k$ |
|   |      |     |     |     |
| d | $-e$ | 0$i$ | 0$k$ | 1$j$ |
|   | $+s$ | 0$j$ | 1$i$ | 0$i$ |
|   | $+w$ | 0$k$ | 0$j$ | 1$k$ |
|   |      |     |     |     |

E

|   |     | B  | G  | R  |
|---|-----|----|----|----|
| u | +*e* | 1*j* | 1*j* | 0*j* |
|   | +*s* | 0*k* | 0*i* | 1*i* |
|   | +*w* | 0*i* | 0*k* | 1*k* |
|   |     |    |    |    |
| d | −*e* | 0*i* | 0*k* | 1*j* |
|   | +*s* | 0*j* | 0*i* | 1*i* |
|   | +*w* | 0*k* | 0*j* | 1*k* |
|   |     |    |    |    |

Here, we define a unit charge as some combination, such as ±*i*s, ±*j*e, ±*k*w, where the individual values of *s*, *e*, and *w* may be 1 or 0. We assume that the charge components *s*, *e* and *w* are fixed with respect to each other and have rotational asymmetry, whereas the quaternion components *i*, *j* and *k* are variable and have rotational symmetry. However, to prevent the charge component *e*, say, being associated with *i* or *k* as easily as it is associated with *j*, we assume that we must always assign unit values of *e* to the term *je* and zero values of *e* to the terms *ie* and *ke*, in such a way that physical systems with 1*je* are indistinguishable from those with 0*ie* and 0*ke*. While this would be impossible if unit charges existed independently, it would be possible if unit charges could only exist in some form of combination, as we observe with the experimentally-discovered mesons and baryons. Individual charges could then be identified but only in such a way as never to be separable.

The tables appear to be the only ways in which charge can be 'accommodated' to a quaternion scheme, while preserving separate conservation laws for each type of charge. The tables are a model, derived inductively, to duplicate the facts of the Standard Model with the minimum of assumptions. We can use the fact that the charges are irrotational, but the quaternions are not, to derive the essential features of the Model. Even in this case, E appears to be excluded by requiring all three quaternions to be attached to specified charges (losing the three required degrees of freedom, and, at the same time, necessarily violating Pauli exclusion[13]). If applied to known fermions, it would appear that A-C must represent the coloured quark system, with *s* pictured as being 'exchanged' between the three states (although, of course, in reality, all the states exist at once), while D-E, with the exclusion of the *s* charge, represent leptons. The antifermions are generated by reversing all charge states, while two further generations are required by the exclusion of negative values of *w* in fermion states by the respective violations of parity and time reversal symmetry.

To read the tables, it is easiest to look at the *d* quarks first, that is to do the charges of a single sign. In principle, we have to have a single sign for any charge type to avoid specifying its quaternion operator in a baryon, thus suggesting that charges of the opposite signs represent antiparticle states. However, the particle-antiparticle option is only available once (like the 'privileged' triad in the Dirac

operator[1]) so, we have to look for other options. Taking the default position as filled instead of empty (+*e* +*e* +*e*), and remembering that (for purely historical reasons) the – sign is the 'normal' one for *e*, we then have the two 'weak isospin' states (+*e* +*e* 0) and (0 0 –*e*). This creates variation of sign without changing the 1 in 3 structure which we suppose to be relevant. For the weak case, we take the option inherent in Dirac's interpretation of –*E* for antiparticles – a filled weak vacuum and consequent charge conjugation symmetry violation. The sign is then immaterial, only the particle-antiparticle nature.

The other representation is algebraic. The tables can be derived by trial and error but we can also develop algebraic models for them, with the rotational elements modelled by vectors. The charge accommodation algebra for quarks can be expressed in the following assignments:

| | |
|---|---|
| down | $-j\mathbf{r}_1 + i\mathbf{r}_2 + k\mathbf{r}_3$ |
| up | $-j(\mathbf{r}_1 - \mathbf{1}) + i\mathbf{r}_2 + k\mathbf{r}_3$ |
| strange | $-j\mathbf{r}_1 + i\mathbf{r}_2 + z_P k\mathbf{r}_3$ |
| charmed | $-j(\mathbf{r}_1 - \mathbf{1}) + i\mathbf{r}_2 + z_P k\mathbf{r}_3$ |
| bottom | $-j\mathbf{r}_1 + i\mathbf{r}_2 + z_T k\mathbf{r}_3$ |
| top | $-j(\mathbf{r}_1 - \mathbf{1}) + i\mathbf{r}_2 + z_T k\mathbf{r}_3$ |

For the corresponding leptons (where $\mathbf{r}_1 = \mathbf{r}_3$ and there is no $\mathbf{r}_2$ term), we have:

| | |
|---|---|
| electron | $-j\mathbf{r}_1 + k\mathbf{r}_1$ |
| *e* neutrino | $-j(\mathbf{r}_1 - \mathbf{1}) + k\mathbf{r}_1$ |
| muon | $-j\mathbf{r}_1 + z_P k\mathbf{r}_3$ |
| *µ* neutrino | $-j(\mathbf{r}_1 - \mathbf{1}) + z_P k\mathbf{r}_1$ |
| tau | $-j\mathbf{r}_1 + z_T k\mathbf{r}_1$ |
| *τ* neutrino | $-j(\mathbf{r}_1 - \mathbf{1}) + z_T k\mathbf{r}_1$ |

$\mathbf{r}_1$, $\mathbf{r}_2$, $\mathbf{r}_3$ are unit vectors, randomly taking the values **i**, **j**, or **k**. These vectors are, in principle, fully independent unit vectors (**i**, **j**, **k**) in a (real) 3-dimensional space. The **1** in ($\mathbf{r}_1 - \mathbf{1}$) is a unit vector. *z* represents charge conjugation violation and has two forms depending on whether it comes with *P* or *T* violation. *z* is not an algebraic operator. It is just a symbol to say that, in treating the *w* of the second and third generator as though it were positive in the same way as the *w* of the first generation, we have to violate charge conjugation symmetry. The two forms are the two complementary violations – parity and time reversal – that go with the charge conjugation. As with the tables, –*j* represents electric charge (traditionally negative), *i* strong, *k* weak. Each term is successively scalar multiplied by the unit vectors **i**, **j**, and **k** to produce the component 'coloured' quarks of the composite baryons; each of **i**, **j**, **k** representing one colour of quark. The antiparticles simply reverse all the signs.

It is important that, even though $\mathbf{r}_1$, $\mathbf{r}_2$, and $\mathbf{r}_3$ are completely random, the number of different outcomes is reduced by repetitions, and is five, as in the anticommuting pentads of the Dirac algebra, rather than, say, 27, and effectively we 'privilege' one of $\mathbf{r}_1$, $\mathbf{r}_2$, $\mathbf{r}_3$ by allowing it complete variation with respect to the others ($\mathbf{r}_2$ being the one selected). This is effectively the same as 'privileging' **p** as a vector term with full variation in the Dirac anticommuting pentad. 27 degrees of freedom are thus reduced to 5 tables because, though $\mathbf{r}_1$, $\mathbf{r}_2$ and $\mathbf{r}_3$ are independent, in principle, it is only the final pattern of 1s and 0s that counts, and, many of the possible sets of $\mathbf{r}_1$, $\mathbf{r}_2$ and $\mathbf{r}_3$ are repetitions, which produce identical patterns of 1 and 0.

The charge conjugation represented by $z$ is brought about by a filled weak vacuum; the terms ($\mathbf{r}_1 - \mathbf{1}$) and $\mathbf{r}_1$, which represent the two states of weak isospin (the −1, of course, really represents +1 if *j* is conventionally negative), are associated with this idea. In a sense the **1** is a 'filled' state, while 0 is an unfilled state. We are, thus, creating two possible vacuum states to allow variation of the sign of electric charge by weak isospin, and linking this variation to the filling of the vacuum which occurs in the weak interaction. The weak and electric interactions are linked by this filled vacuum in the $SU(2)_L \times U(1)$ model, as they are here by our description of weak isospin, and the $SU(2)_L$ comes from the two states of weak isospin in the charge-conjugation violated (hence left-handed) case.

It seems that, by privileging $\mathbf{r}_2$, as described here, we can reduce the algebra to something very similar to the Dirac algebra, though we use a commuting, rather than an anticommuting, set of elements. In our previous work,[1] we defined the creation of the Dirac state as a process of 'compactification' of the eight basic units of the algebra (*i*, **i**, **j**, **k**, 1, *i*, *j*, *k*) into a more 'primitive' anticommuting pentad (*i***k**, *i***i**, *i***j**, *i***k**, *j*) (which must always take this, or a similar form[14]). If we take the quarks only and confine ourselves to the *d* representation, the quark charges can be represented by a pentad of the form: *j***j**, *i***i**, *i***j**, *i***k**, *k***k**, (for –*e*, *s*, *w*, respectively). This pentad can generate the whole of the Dirac algebra, but it is not an anticommuting pentad: *j***j**, *i***i**, *k***k** commute, but *i***i**, *i***j**, *i***k** anticommute. Also, the full range of + and − terms is not generated unless all the terms anticommute with each other. The pentad is not a nilpotent, and presumably requires a scalar term, for which there is an obvious candidate in mass, which is actually a symmetry-breaking term.

Symmetry-breaking is, in fact, a clear consequence, of the setting up of this algebraic model. The charge accommodation procedure does not come directly from 'compactification' in the same way as *E*-**p**-*m*, though it is actually a result of this 'compactification'. We assign quaternions to $\mathbf{r}_1$, $\mathbf{r}_2$, $\mathbf{r}_3$, from the fact that we have quaternion charges, and so create the Dirac state, but the charges (though individually conserved as physical entities) cannot be assigned uniquely to these quaternions, and we create the appropriate extra degrees of freedom by having the random vector components $\mathbf{r}_1$, $\mathbf{r}_2$, $\mathbf{r}_3$ attached, and only removing these by scalar producting with **i** + **j** + **k** in the combined baryon state. However, the algebra of such combinations requires compactification into a more fundamental pentad, which privileges one of the charges as retaining its full degrees of freedom. We happen to call this charge *s*.

The algebraic treatment of charge accommodation also helps to make sense of the various strange physical consequences of the combined electroweak interaction.

As we have seen, the random vector quantities are completely random, but repetitions reduce the number of real options in producing the tables. So, when time, space and mass map onto the charges *w*-*s*-*e*, only one of the charges (*s*) has the full range of vector options, as it is only the relative values of the vectors that count. If we take the *d* quarks as being the 'standard', *s* charges have the full vector variation. If we 'fix' one of the others (say *e*) for *s* to vary against, then there are only 2 remaining options for *w*, unit on the same colour as *e* or unit on a different one. We can refer to this as *w* 'on' and 'off' *e*.

In fact, if the full variation of *s* is to be allowed, and the combination of *w*, *s*, *e* all 'on' is forbidden, then the combination of *w* and *e* both 'on' can only happen in the absence of *s*. (This creates the lepton states for D/E, as opposed to the quark states for A/B/C.) The reason why we fix *e*, of course, rather than *w*, is because the mapping has made *e* mass-like, and *w* time-like. The time-like *w* has two mathematical states (like *T* or *E*), while *e* has one (like *m*).

The weak interaction can be thought of as a swapping of *w* from *e* 'on' to *e* 'off' or vice versa, creating the $SU(2)_L$, but, in fact, there is no mechanism for doing this directly, as there is in the strong interaction, because there is no combined system to do it in. What we *can* do, however, is to annihilate and create, and instead of swapping over *w*, we annihilate and create *e*, either filling the vacuum or emptying it. However, we cannot annihilate or create a charge without also annihilating or creating its antistate, and the weak interaction (unlike the strong) always involves the equivalent of particle + antiparticle = particle + antiparticle, or a double particle interaction going both ways at once. We don't know which it *really* is because the weak interaction works to *prevent* such knowledge.

It is because of the filling and emptying of the vacuum via the *e* charge that rest mass is involved. $W^+$ and $W^-$ involve a one-way *e* transition, $Z^0$ involves a two-way *e* transition, the purely electromagnetic ($U(1)$) $\gamma$ no *e* transition. This gives us 0.25 for the electric / weak ratio. The same value also occurs for the weak isospin quantum number squared. Weak isospin is this annihilation and creation, creating a vertical motion in the tables. Sideways motion (swapping over *w*) never happens directly, but can be considered to happen indirectly, and where the signs are those for the antiparticle, this will be taken care of by the particle and antiparticle interactions being always simultaneous, with the weak interaction unable to recognise the difference. Particle + antiparticle also allows for the 'elimination' of *s* in those transitions, like neutron beta decay, that appear to be from quark to lepton. In transitions that appear to be A/B/C to D/E (neutron decay) the interaction can be represented by two vertical transitions acting in opposite directions.

Overall, the asymmetry between the interactions is a matter of what (algebraic) options have been 'used up', and this relates directly to the formation of an $SU(5)$ / $U(5)$ algebra for Grand Unification.[1] There is only one 'privileged' vector option, and once this has been used, then variation becomes very limited. The set of matrices for the *w* transitions effectively reduce the eight $SU(3)$ ones to four $SU(2)_L \times U(1)$ ones, two of which are identity transformations. This is decided on the basis that all the alternative ones are either forbidden or taken up by those of *s*. The exclusively left-handed aspect of the weak *SU*(2) occurs because we have no remaining options for varying the sign of *w*, once we have decided that we have to eliminate *s* in the *w* 'on' state.

**10 The Dirac algebra and charge accommodation**

From the results outlined in sections 5-7, it would appear that the vectors (and the scalar products) involved in the algebraic representation for charge accommodation have a real physical meaning, although they were originally introduced as a convenient formal device. In charge accommodation, we have the vectors **r**$_1$, **r**$_2$, **r**$_3$ (representing random units of **i**, **j**, or **k**), which provide the extra degrees of freedom needed to apply conservation rules to charge at the same time as we apply 'nonconserved' quaternion labels. The vector element is then removed by taking the scalar product with a full unit vector (**1**). The strong charge (*s*), for example, cycles (in a gauge invariant way) through the values **i**, **j**, **k**. When we apply the concept to the confining force for the quarks in a baryon, we have an angular momentum term, σ.**p**, cycling (in an equally gauge invariant way) through the possible orientations of **p**, with σ.**p** having exactly the same form as, say, **r**$_2$.**1**.

Ultimately, it is the angular momentum term (**p** or σ.**p**), which carries the information concerning charge conservation, and the three charges are separately conserved because they represent three aspects of the angular momentum conservation process. The random vectors represent angular momentum states, even when associated with weak and electric charges, where they are associated respectively with the sign, and the magnitude, through the connections of **p** with *E* and **p** with *m*. In these cases, we are not concerned with the directional components, which is entirely associated with the strong charge, and this is responsible for the fact that we are able to associate a fixed single vector (**i**, **j**, or **k**, though the choice is arbitrary) with each of the quaternion labels (***k*** and ***j***) specifying *w* and *e*.

In the Dirac state vector, an angular momentum state must remain unspecified as to direction, although one direction (and one direction only) may be well defined. There are consequently two ways of constructing a fermion wavefunction. One specifies the three components of the angular momentum, and allows the coexistence of three directional states as long as none is specified. This 'baryon' structure requires a quaternion state vector of the form:

$$(kE \pm ii\, p_1 + ij\, m)\, (kE \pm ii\, p_2 + ij\, m)\, (kE \pm ii\, p_3 + ij\, m)\,,$$

with six coexisting representations for the 'three colour' or 'three quark' combinations. The other, 'free fermion', structure specifies the total angular momentum, and has a quaternion state vector of the form:

$$(kE \pm ii\, \mathbf{p} + ij\, m)\,,$$

which is well defined in a single direction (though without specific preference).

The first type incorporates the three specified directional components by specifically requiring the strong charge (the name associated with the *i* quaternion label) to be cycled in the manner specified by **r₂.1**. A consequence of this is that spin is not intrinsic to the quarks but is a property of the system. A spin direction is uniquely definable only for the baryon as a whole, and not for the component quarks. This becomes significant when we investigate the behaviour of the vectors assigned to the other two charges, because we must assume that they are not aligned. The second type of state vector (the free fermion or lepton) must necessarily exclude the strong charge or intrinsically directional components of angular momentum. The angular momentum must have a single well-defined direction, and so the random vectors associated with the electric and weak charges must be aligned. In fact, alignment of these vectors can be taken as the signature of a free fermion, excluding the strong interaction.

Like those of the strong charge, the conservation properties of the weak and electromagnetic charges are determined by those of the angular momentum operator. However, neither of these is attached directly to **p**; one is attached to *E* and one to *m*, and it is the *combination* of these which affects **p**. It is for this reason that we think of the electric and weak forces as being in some way combined. In principle, the charge represented by the quaternion label *k* (which we call the weak charge) produces two sign options for *iE*, because the algebra demands complexification of *E*, and there are necessarily two mathematical solutions. Only the positive solution, however, should be physically meaningful, and we compensate by creating a filled vacuum for the ground state of the universe, in which states with negative *E* (or antifermions) would not exist, though they are allowed by the parallel mathematical status of the quaternion labels as square roots of –1, which permits charge conjugation or reversal of the signs of the quaternion labels.

The result of the filled (*k* or weak) vacuum is the violation of charge conjugation symmetry for the weak interaction, with consequent violation of either time reversal symmetry or parity. One manifestation of these violations is that both the *w* and *s* charges takes only one effective sign for fermions, though charge conjugation should allow two signs for *w* if that of *s* is fixed. The suppression of the alternative sign for *w* (which we arbitrarily designate as –) means that quark and free fermion states become mixed states, for at least one of the isospin options, containing +*w*, and suppressed –*w* states involving respective violations of parity and time reversal symmetry. Another manifestation is that only one state of σ.**p** exists for the pure *w* interaction for fermions – and, because σ = –**1**, this is the state of negative helicity or left-handedness. To create the states of positive helicity or right-handedness, which should also exist for –**p**, we have to introduce mass, which is associated in the Dirac state with the *j* quaternion label, which defines what we call the electric charge. The introduction of *m* also introduces the *E* / **p** mixing, which produces a right-handed component mixed with the left-handed. Such a mixing can only be produced by a mixing of the effects of *e* charges with those of *w*.

The presence or absence of *e* charges creates the characteristic $SU(2)_L$ 'isospin' pattern associated with the weak interaction, for this interaction must be both uniquely left-handed for fermion states and indifferent to the presence or absence of the electric charge, which introduces the right-handed element. The $SU(2)$ produces a quantum number, $t_3$, such that $(t_3)^2 = (½)^2$ in half the total number of possible states. For free fermions, with 0 or ±1 as the quantum number for the electric force, and so with $Q^2 = 1$ in half the total number of possible states, the key electroweak mixing parameter becomes $\sin^2 \theta_W = \Sigma (t_3)^2 / \Sigma Q^2 = 0.25$, which is the same proportion as would be obtained by taking the electron and neutrino as the possible free fermion states. Since the weak force must also be indifferent to the presence of the strong interaction, or to the directional state of the angular momentum operator, then the same mixing proportion must exist also for quark states, and separately for each colour, so none is preferred.

For the weak and electric forces to carry no directional information, the charges and their associated vectors must be arranged for only one of the three quarks in a baryon to be differentiated at any instant, and the *e* and *w* values so specified must be separated. Thus if we define the weak isospin states for the specified colour for *e* as 0 and –*e* (the negative value being adopted by convention), then the only corresponding isospin states for the other colours that retain both the accepted value of $\sin^2 \theta_W$ and the variation of only one quark in three, are *e* and 0. In effect, this is like adding a full *e e e* background or 'vacuum' to the original 0 0 –*e*, so that the two states of weak isospin in the three colours become:

| *e* | *e* | 0 |
| 0 | 0 | –*e* . |

The creation of three generations, as well as isospin states, results from the violations of parity and time reversal symmetry which are consequent upon the effective suppression of –*w* states for fermions.

The charge accommodation rules can thus be derived entirely from the Dirac representation. The idea that rotationally conserved charges should have a further degree of freedom to allow for variation of the quaternion labels is paralleled by the use in the Dirac representation of a real (axial) vector quantity, angular momentum, whose conservation is equivalent to an invariance to spatial rotation. The introduction of a scalar product of the vector with σ, in the quantized Dirac representation, is equivalent to 'quantizing' the charge accommodation algebra (to 0 or 1) by taking the scalar product with the unit vector **1**. The variation of only one quark in three for each charge is a necessary consequence of the use of a spherical rotational system for the variation. The quark tables A-C are consequences of the necessary nonalignment of the *w* and *e* charges in the baryon state; while the lepton states D-E, excluding *s*, are a necessary consequence of an alignment. The pattern of *e* charges in the quark system is a necessary consequence of maintaining the weak charge's indifference to the presence of both *s* and *e*.

## 11 Spin and statistics

In previous work, one of us predicted on symmetry grounds that an extension of Noether's theorem would require the conservation of the type of charge (*w*, *s*, *e*) in fermion or boson states to be exactly equivalent to the conservation of the states' angular momentum.[15] The reason for this theorem now becomes apparent. The conservation of quantized angular momentum incorporates three separate conservation laws for boson or fermion states: the conservation of the single well-defined direction, irrespective of the component contributions; the conservation of orientation (up or down); and the conservation of magnitude. The first requires the conservation of strong charge, the second conservation of weak charge, and the third conservation of electric charge. Each is independent of the other, and so conservation of angular momentum requires the separate conservation of each of the individual charge components, though none of the charges represents angular momentum as such (or, indeed, energy or mass as separate quantities).

Related to this conservation principle is the spin-statistics theorem, which associates bosons to states with integral units of spin (± 1 or 0) and fermions to states with half-integral units (spin ± ½ or ± 3/2). We have already shown that the origin of these values lies in the mathematical procedure which makes a fermion QSV effectively the 'square-root' of the QSV for a boson, and that the process has analogies with the concepts of scalar and vector addition.[1] They can also be related to the presence or absence of particular charges in fermion and boson states. In our representation, leptonic fermions have unit weak charge (± *kw*); bosons have zero weak charge. Essentially, to create the fermion state, we have a nonzero *k* term in the Dirac equation, because the complex term *ikE*, together with at least one noncomplex term in *i* or *j*, is essential to create a nilpotent anticommuting operator; physically, this is the same as saying that energy is necessary, along with at least one of momentum or mass. The spin ± ½ term is thus an indicator of the presence of *w*.

In the baryon system, composed of three fermion-like states, which may be conveniently described as 'quark fermions',[14] this manifests itself in the form of a spin ± ½ or ± 3/2 for the *combination*, being effectively 'transferred' between the components. The gauge invariance of the process is maintained by the gauge invariance of the 'transfer' of the strong charge *s*. Thus *s*, in the baryon state, is a measure of the spin ± ½ or ± 3/2 angular momentum, which is independent of that of *w*. The conservation of the spin angular momentum requires the separate conservations of *w* and *s*. In principle, also, this requires the separate conservation of *e*, even though the *e* value does not determine the spin quantum number of the fermion, because if *w* and *s* are separately conserved, then they cannot be converted into *e*; the conservation of *e*, also, according to the London argument of 1927, relates to the conservation of the linear momentum **p**, which determines the *value* of the spin angular momentum in an applied electric field, and also the conserved orbital angular momentum, which this field produces. So, the separate conservation of the three charges is directly experienced in the total conservation of the angular momentum within a system in which they operate. By analogy with angular momentum, in which one direction, at any instant, remains uniquely defined, the conservation law may be defined in terms of any one 'direction' of the 3-dimensional charge parameter.

The angular momentum state of a boson, of course, is derived from the combined angular momenta of its component fermion and antifermion states, and so will necessarily be specified by integral quantum numbers. It is significant that the fermion state represented by a QSV of the form ($\pm k E \pm i i \mathbf{p} + i j m$) derives its noncommutativity, and hence its 'fermionic' characteristics, from the 'hidden' presence of the charges represented by the quaternion labels $k$, $i$ and $j$.

**Appendix: Summary of procedures used in generating the Dirac formalism**

The mathematical details are given in our earlier work,[1] but the steps in the procedure might be conveniently described as follows:

(1) On the assumption that mass-charge is described by a quaternion, while space-time is described by a 4-vector, create a 32-part algebra from their combination. The $i$ term of the 4-vector ensures that this is a complexified 4-vector-algebra.

(2) Find a primitive set of components which will generate the whole algebra that will match the gamma matrices of the Dirac algebra. This turns out to be an anti-commuting pentad, but it is significant that it can exist in more than one form.

(3) Invoke the presumed full symmetry between space-time and mass-charge to make the 4-vector 'quaternion-like', which means that the vector part becomes multivariate, or isomorphic to the set of Pauli matrices. Multivariate vectors have a 'full' product that is equivalent to the scalar product plus $i$ times the vector product.

(4) Use existing results involving multivariate vectors applied to the momentum operator in the Schrödinger equation[2] to hypothesize that spin is generated by the multivariate nature of this operator, and does not require *extra* spinor terms in the wavefunction. The spin here comes from the extra vector product term in the full product.

(5) Apply the first four terms of the pentad (equivalent to $\gamma^0$, $\gamma^1$, $\gamma^2$, $\gamma^3$) to the standard form of the Dirac equation to produce a vector-quaternion form of the differential operator, with an unspecified wavefunction.

(6) Multiply this equation throughout from the left by $\gamma^5$ or $ij$.

(7) Recognize that one can redefine the $\gamma^0$, $\gamma^1$, $\gamma^2$, $\gamma^3$ terms using the variation allowed within the anticommuting pentads to create a new form of the Dirac equation in which the mass term is preceded by $ij$. In this form, the differential operator, when reduced to eigenvalues, becomes a nilpotent of the form ($kE + ii \mathbf{p} + ij m$).

(8) Recognize that, if this equation is valid, and a plane wave solution is applied, then the wavefunction must also be a nilpotent incorporating the term ($kE + i\boldsymbol{i}\ \mathbf{p} + i\boldsymbol{j}\ m$).

(9) Recognize that four solutions are immediately made possible by the use of $\pm E$ and $\pm \mathbf{p}$, with the multivariate nature of $\mathbf{p}$ allowing us to interpret it as either $\mathbf{p}$ or $\sigma.\mathbf{p}$ (if we can show that the correct value of spin is generated from our equation).

(10) Use the classical energy-momentum-rest mass equation to derive simultaneously all four solutions by the same quantization procedure of replacing eigenvalues by differential operators, as for the Schrödinger equation.

(11) Recognize that this derivation allows us to describe the variation in the four solutions *either* by varying the exponential term or by varying the differential operator (but not both). (This is related to the Feynman description of negative energy particles being equivalent to positive energy particles going backwards in time, with reversal of the spin / momentum terms being equivalent to reversal in position coordinate.)

(12) Demonstrate that a matrix version of the equation allows us to vary the differential operator (and that we can even reduce to a single differential operator, if we construct our matrices in a particular way).

(13) Recognize that varying the differential operator rather than the exponential, and incorporating the spin concept into a multivariate $\mathbf{p}$, allows us to construct a wavefunction made up of a column vector with terms ($kE + i\boldsymbol{i}\ \mathbf{p} + i\boldsymbol{j}\ m$), ($kE - i\boldsymbol{i}\ \mathbf{p} + i\boldsymbol{j}\ m$), ($-kE + i\boldsymbol{i}\ \mathbf{p} + i\boldsymbol{j}\ m$), ($-kE - i\boldsymbol{i}\ \mathbf{p} + i\boldsymbol{j}\ m$), and a *single* exponential term. This liberates the wavefunction from being confined to being an ideal, although a nilpotent could be taken as being an extreme case of an ideal.

(14) Recognize that, since the wavefunction operator and the eigenvalue of the differential operator are in every case identical, then we can assume that $\mathbf{p}$ means $\mathbf{p}$, $\sigma.\mathbf{p}$, in general or in any specified direction, without loss of generality, and that it can be for a field-free particle or a $\mathbf{p}$ (or $E$) involving field terms.

(15) Recognize that, if this description is valid, we can now generate all required wavefunctions using a purely operator approach. Everything emerges from using a bra vector with four operator terms, or a ket vector with four operator terms, or combinations of these.

(16) Demonstrate that this form of the equation produces the correct spin and helicity relations.

(17) Demonstrate that one can start from this form of the equation, and, by various transformations, arrive at the standard Dirac equation.

(18) Demonstrate that one can start from the Dirac equation in any of its forms, and arrive at this new form of the equation. Inevitably, such derivations will always require some specification of the representation, since the two equations are *physically equivalent*, not isomorphic, but, in any individual case, such a representation can always be made.

(19) Define the procedure for normalisation in the form $\psi\psi^*$.

(20) Using the adjoint wavefunction, derive the bilinear covariants to show that the Dirac current is zero in the absence of an external field, and construct the appropriate Lagrangian.

(21) Construct annihilation and creation operators, and show that second quantization is unnecessary since the new form of the equation can be shown to be derivable from the quantum field integrals.

(22) Demonstrate that routine results, such as the hydrogen atom, follow just as easily from this model as from any other.

From these steps, we believe that we can produce the following, most of which are described in our earlier work:[1]

(1) A vacuum wavefunction, of which each nonzero term is $k$ times the terms of the state vector and exponential term for the fermion, with a vacuum *operator*, of which each nonzero term is just $k$ times the terms of the state vector. All fermion states which may be produced may then be considered as acting on the vacuum wavefunction, and the exponential part of the fermion wavefunction be regarded as, in origin, a vacuum term, expressing all possible space and time variations of a state in the vacuum.

(2) C, P and T transformations using the three quaternions operators, with consequent demonstration of CPT symmetry, etc.

(3) Immediate Pauli exclusion for identical fermions.

(4) A clear reason why the Dirac equation cannot apply to bosons, although the Klein-Gordon equation can be applied to fermions. A nilpotent wavefunction requires a differential operator with nilpotent eigenvalue to produce a zero product.

(5) An operator for the antifermion that reverses the $E$ signs in the fermion operator, and that can be conveniently arranged as a bra vector if the fermion operator is a ket, and vice versa.

(6) Production of boson states by allowing a bra antifermion operator to act on a ket fermion state. The product is always a scalar.

(7) Differentiation between vector and scalar bosons according to whether the signs of the fermion and antifermion **p** terms are the same or different.

(8) An explanation of why vector bosons may be massless, but scalar bosons may not.

(9) An expression for the scalar wavefunction of a Bose-Einstein condensate, made of two fermions with opposite **p** states.

(10) An expression for the wavefunction of a baryon, with three symmetric and three antisymmetric terms, as required.

(11) Correct parity values for ground state baryons and bosons.

(12) A connection with approaches to the Dirac algebra that are constructed using quaternion or 4-vector operators to represent the four solutions, and an *explanation* for the existence of four solutions, and the validity of a $4 \times 4$ matrix for the differential operator.

(13) Annihilation and creation operators for the quantum field that are identical to the nilpotent operators and have the required algebraic relations. Our formalism already thus incorporates the quantum field, and can be shown equivalent to it.

(14) Supersymmetry operators that are identical to the bra and ket vectors used for fermions and antifermions.

(15) Infinite vacuum operation by a fermion state which is identical to an infinite alternating series of virtual boson and fermion states, as required for renormalization. Similar vacuum operation by bosons.

(16) Propagator terms for QED that immediately relate the Dirac propagator to the Klein-Gordon propagator, using the bra or ket operator for the fermion, and not requiring any averaging over helicity states.

(17) Explanation of the *origin* of the Dirac state, in reducing the eight original components of the algebra $(1, i, j, k, i, \mathbf{i}, \mathbf{j}, \mathbf{k})$, derived from the physical mass, charge, time, and space, to a Dirac pentad, in which mass, time and space acquire, by combination, the characteristics of the three 'dimensions' of charge, when this disappears as an independent entity.

(18) The consequent origin of quantized rest mass in the Dirac state.

Our approach eliminates much of the matrix algebra usually associated with the Dirac equation, and allows all tasks to be performed with a single quaternion vector operator. No mysterious wavefunctions or spinors are hidden in the formalism. In addition, we believe that the group symmetry for the Dirac algebra is, in principle, identical or closely related to that for the Standard Model, and that the relations between the two formalisms will be more clearly revealed by basing the Dirac formalism on fundamental physical symmetries, rather than algebraic generalities.